\documentclass[final,5p,times,twocolumn,authoryear]{elsarticle}

\usepackage{amssymb}
\usepackage{amsmath}

\DeclareUnicodeCharacter{2212}{\textminus}

\usepackage{stfloats}
\usepackage{float}
\usepackage[T1]{fontenc}
\usepackage[utf8]{inputenc}
\usepackage{url}
\usepackage{textcomp}

\usepackage{booktabs}
\usepackage{caption}
\usepackage{tabularx}
\usepackage{threeparttable}
\usepackage{makecell}
\usepackage{array,ragged2e,hyphenat}

\newcolumntype{Y}{>{\RaggedRight\arraybackslash}X}

\journal{Computers in Human Behavior Reports}

\begin{document}

\begin{frontmatter}


\author[inst1]{David Stammer\corref{cor1}\fnref{fn1}}
\ead{david.stammer@tuwien.ac.at}

\author[inst2]{Hannah Strauss}
\ead{hannah.strauss@uibk.ac.at}

\author[inst1]{Peter Knees}
\ead{peter.knees@tuwien.ac.at}

\affiliation[inst1]{organization={TU Wien, Faculty of Informatics},
            addressline={Favoritenstraße 9-11/194-4}, 
           city={Vienna},
            postcode={1040}, 
           country={Austria}}

\affiliation[inst2]{organization={University of Innsbruck, Department of Psychology},
            addressline={Universitätsstraße 15}, 
           city={Innsbruck},
            postcode={6020}, 
           country={Austria}}


\title{Perception of AI-Generated Music: The Role of Composer Identity, Personality Traits, Music Preferences, and Perceived Humanness} 



\begin{abstract}
The rapid rise of AI-generated art has sparked debate about potential biases in how audiences perceive and evaluate such works. This study investigates how composer information and listener characteristics shape the perception of AI-generated music, adopting a mixed-method approach. Using a diverse set of stimuli across various genres from two AI music models, we examine effects of perceived authorship on liking and emotional responses, and explore how attitudes toward AI, personality traits, and music-related variables influence evaluations. We further assess the influence of perceived humanness and analyze open-ended responses to uncover listener criteria for judging AI-generated music. Attitudes toward AI proved to be the best predictor of both liking and emotional intensity of AI-generated music. This quantitative finding was complemented by qualitative themes from our thematic analysis, which identified ethical, cultural, and contextual considerations as important criteria in listeners’ evaluations of AI-generated music. Our results offer a nuanced view of how people experience music created by AI tools and point to key factors and methodological considerations for future research on music perception in human–AI interaction.
\end{abstract}




\end{frontmatter}


\section{Introduction}
\label{sec1}
High-fidelity generative models have brought AI art into the mainstream, challenging traditional notions of creativity \citep{agres_evaluation_2016, wingstrom_redefining_2024} and authenticity \citep{salas_espasa_aura_2025}.
Building on recent advances in text-to-music models \citep{agostinelli_musiclm_2023, copet_simple_2023}, consumer platforms such as \textit{Suno} and \textit{Udio} now enable non-expert users to generate polished audio tracks with minimal input and effort. 

With AI systems increasingly shaping or replacing creative processes, understanding how people perceive and respond to AI-generated art becomes crucial \citep{HONG2022107239}. From an applied AI perspective, assessing user acceptance informs system design and interface decisions; from a cultural standpoint, perceptions can affect the legitimacy of AI-assisted artworks and the economic well-being of professional creators \citep{tatar_shift_2024}. 

Empirical work on AI-generated art reveals complex evaluative dynamics: People tend to devalue AI-generated art and ascribe less creativity to it \citep{di_dio_art_2023, horton_jr_bias_2023, millet_defending_2023}. However, people ascribe emotion and intentionality to AI-generated artworks, regardless of perceived authorship --- whether they believe that the creator is human or AI \citep{demmer_does_2023}. Perceptions about AI-art are influenced by various factors, including contextual information \citep{bara_ai_2025}, pre-existing schemata or attitudes toward AI \citep{HONG_AI_2019}, and beliefs about human creativity \citep{millet_defending_2023}. 
Against this backdrop, we now focus on the characteristics of AI-generated music and how listeners evaluate and perceive it.

\subsection{AI-Generated Music}

AI-generated music refers to music created wholly or in collaboration with algorithmic systems, ranging from fully autonomous models to interactive, human–AI co-creative tools \citep{briot_artificial_2021}. Until 2022, research focused on symbolic music generation, where models produce sequences of musical notes (e.g., MIDI or sheet music) rather than raw audio. The field expanded rapidly, with models covering a range of different music generation tasks, from simple melody generation, to multi-track arrangements, and music generation in different styles, with structure or timbre \citep{miguel_civit_systematic_2022, herremans_functional_2018, ji_survey_2024}. Building on these advancements, co-creative systems allow users to iteratively steer musical ideas, by selecting, editing, or constraining AI-generated suggestions \citep{louie_novice-ai_2020}.
Another line of research explores music generation with emotions with models targeting specific affective states \citep{bao_generating_2023, huang_emotion-based_2020}. 
A major technical leap has been the move from symbolic generation to producing actual audio. Rather than outputting notes for humans or software to perform, newer models can directly synthesize musical waveforms 
\citep{dong_deep_2022}, leading to text-to-music systems such as GoogleLM \citep{agostinelli_musiclm_2023}, MusicGen \citep{copet_simple_2023} or Mousai \citep{schneider_mousai_2023}. These advances underpin commercial platforms like Suno and Udio, which turn natural-language prompts into polished tracks of different genres including lyrics in various languages \citep{choi_understanding_2025}.

In terms of perception of AI-generated music, research in computational creativity and applied machine learning has examined the question of the creative value of the output of these models with subjective listening tests: Listeners rate samples of a proposed music generation model in comparison to different baseline models or human composed music \citep{yang_evaluation_2020}. Frequent evaluation metrics include overall quality, creativity or novelty, as well as music specific metrics assessing melody, harmony, and rhythm of generated music samples \citep{chu_empirical_2022}. Another key metric is the assessment of naturalness or humanness. This is either directly assessed via rating scales or examined using a musical Turing Test \citep{turing_computing_1950}, in which listeners judge whether a piece was created by a human or an AI system, under conditions where they may be informed, uninformed, or intentionally misled about the source \citep{ariza_interrogator_2009}. Within specific tasks, such as the generation of Bach chorales, even early models such as Deepbach \citep{hadjeres_deepbach_2017} achieved results that were difficult to distinguish from original compositions. In order to evaluate emotional aspects of generated music, listeners are typically asked to identify the intended target emotions by categorizing them via dimensional valence-arousal models or with discrete emotions such as "happy" or "sad" ~\citep{dash_ai-based_2023}.

\subsection{AI-music bias}
With AI-generated music becoming increasingly sophisticated, studies have investigated whether listeners evaluate such music differently depending on their beliefs or information about its origin. The empirical evidence, however, is far from uniform:

Early experimental studies manipulating information about composer identity with participants rating live recordings of musicians who performed both human- and computer-composed string-quartet pieces found no significant composer bias \citep{pasquier_investigating_2016}. Similarly, \citet{zlatkov_searching_2023} did not observe a systematic bias against AI-composed music using contemporary pop music samples as stimuli. Consumer marketing oriented research by
\citet{moura_artificial_2021} examining ratings of classical and pop music pieces also found no consistent preference based on authorship.

In contrast, an examination of AI bias across various forms of art by \citet{millet_defending_2023} revealed that listeners felt less awe and attributed less creativity to music pieces they thought were AI-composed. \citet{deguernel_investigating_2022} found evidence of bias among Irish folk music practitioners, indicating that in genres where tradition and authenticity are particularly relevant, negative attitudes toward AI-authorship might be more prevalent. Using a different phrasing, \citet{tubadji_cultural_2021}'s findings suggest a cultural proximity bias \textit{for} human composed music. In a setup in which participants rated the same music pieces both before and after they were informed about the true nature of origin, AI-generated music pieces were rated less favorably. The results also indicated that music experts were more critical about AI-music pieces than non experts. This is in line with the findings of \citet{moura_artificial_2021} with professionals showing lower affective responses than casual listeners, underlining that perception might be linked to personal characteristics.

Beyond individual differences, composer bias has also been shown to vary depending on the musical genre, pointing to contextual factors in how AI authorship is perceived. In a series of studies examining composer bias, \citet{shank_ai_2023} found no effect of composer identity on liking when participants listened to electronic music, which the authors attributed to the genre’s “AI-sounding” qualities. However, when using classical music excerpts, which were generally perceived as more 'human-sounding', AI authorship led to lower liking ratings, suggesting that bias may be genre-dependent. 
Extending \citet{shank_ai_2023}'s approach to performance, \citet{ansani_ai_2024} found that participants rated identical piano performances more positively when they believed the performer was human rather than AI. These effects were independent of musical expertise but moderated by attitudes toward AI. 

Such findings highlight the role of listeners’ pre-existing attitudes and expectations. For instance, \citet{hong_2021_artificialmozart} showed that a more favorable attitude toward creative AI was associated with higher music ratings, in a study focused on people's expectancies of AI music in different genres. Expectancy violations negatively affected the evaluations, although no significant genre differences emerged.

\subsubsection{Perceived humanness of AI-generated music}

\looseness -1
Another central question in people's evaluation of AI-generated music is whether the output is perceived as human-composed. 
In the context of this study, it is of importance which aspects of music lead to participants' attributions as being human or AI. According to \citet{shank_ai_2023}, genre is one aspect, with electronic music more likely to be attributed to an AI composer.  Interestingly, when mislead about the true origin of music,  participants reported perceived musical differences that did not exist, suggesting that expectations and beliefs about humanness shape aesthetic judgment \citep{ansani_ai_2024}.
Qualitative analysis by \cite{deguernel_investigating_2022} revealed evaluation strategies or criteria based on musical characteristics: Participants primarily identified AI-composed tunes by noting excessive repetition, rigid or unnatural structure, and atypical harmonic or melodic progressions. In contrast, human compositions were associated with melodic variation, ornamentation, stylistic familiarity, and features that suggested instrumental playability.
These associations reflect a broader pattern in which musical features perceived as “human” are evaluated more positively, while those linked to AI are viewed less favorably. In line with this, \cite{HONG2022107239} investigated how perceived anthropomorphism and creative autonomy shape the acceptance of AI as musicians. Participants evaluated AI-generated pieces from multiple genres and rated the music’s quality alongside their attitudes toward AI agents. The results showed that higher anthropomorphism and greater perceived autonomy were associated with increased acceptance of AI musicians.
Furthermore, \cite{sun_would_2023} showed that anthropomorphism affects perceived competence and warmth of AI musicians, both of which contribute to more positive attitudes.

\section{Research questions}
Building on the limited and at times inconsistent evidence from previous research, the present study aims at examining the effect of composer identity, genre, and listener characteristics in more detail. Drawing on a diverse set of music stimuli generated by two distinct text-to-music systems, we address four overarching research questions.

\noindent \textit{RQ1: How does information about the composer’s identity influence listeners’ liking and emotional perception of AI-generated music? }

Drawing on previous findings on AI composer bias \citep{moura_artificial_2021, shank_ai_2023, tubadji_cultural_2021}, we investigate whether participants evaluate music less favorably when they believe it was composed by an AI. Unlike previous studies using human-composed stimuli \citep[e.g.,][]{shank_ai_2023} or a small number of songs for comparison \citep[e.g.,][]{hong_2021_artificialmozart, HONG2022107239, moura_artificial_2021, tubadji_cultural_2021}, we use a wide range of stimuli for different genres and two different AI music models. We extend our analysis of composer bias on nuanced assessments of emotional music perceptions, using a domain-specific model. Additionally, we investigate the influence of genres on music perception, extending previous approaches \citep[e.g.,][]{hong_2021_artificialmozart, HONG2022107239} with a variety of 15 different (sub-)genres. 

\vspace{0.2cm}
\noindent H1: AI-generated music is liked less when participants are informed that it was AI-generated
\vspace{0.2cm}

\noindent H2: Perceived emotional intensity of music is lower when participants are informed that it was AI-generated
\vspace{0.2cm}

\noindent H3: Biases for liking and emotional assessment differ depending on music genre. 
\vspace{0.2cm}

\noindent \textit{RQ2: How do attitudes toward AI, individual engagement with music, and personality influence liking and emotional perception of AI-generated music?}

While previous research has identified listener's attitudes toward AI as a central factor in shaping responses to AI-generated music \citep{ansani_ai_2024, hong_2021_artificialmozart}, other characteristics relevant to the emotional experience of music \citep{strauss_emotion--music_2024} have remained largely unexplored. To obtain a more comprehensive listener profile, we hence not only consider attitudes toward AI \citep{stein_attitudes_2024}, but also examine the influence of personality traits \citep[Big-5;][]{mccrae_five-factor_2008}, music competence and appreciation \citep{zentner_assessing_2017, zentner_encapsulating_nodate}, motives for music listening \citep{chamorro-premuzic_personality_2007}, and music genre preferences \citep{rentfrow_re_2003}. Accordingly, we hypothesize that:

\vspace{0.2cm}
\noindent H4: More positive attitudes toward AI are associated with greater liking and stronger emotional responses to AI-generated music. 
\vspace{0.2cm}

\noindent H5: Individual differences in personality, music preferences, listening motives, and musical expertise account for additional variance in liking and emotional responses of AI-generated music.
\vspace{0.2cm}

\noindent \textit{RQ3: How does perceived humanness of AI-generated music influence liking and emotional perception?}

An important aspect in the evaluation of AI-generated music is the assessment of naturalness or humanness. Previous approaches \citep{ansani_ai_2024, shank_ai_2023} used an experimental design with human composed music pieces, misleading participants to believe that they were AI-generated. We investigate whether perceptions of humanness are linked to higher liking and stronger emotional responses, using actual AI-generated music pieces. In addition, we explore which aspects of AI-generated music (in terms of genre) lead to higher humanness ratings.

\vspace{0.2cm}
\noindent H6: AI-generated music is liked more when it is perceived as human-created, regardless of composer-identity. 
\vspace{0.2cm}

\noindent H7: Higher perceived humanness is associated with stronger emotional intensity ratings.
\vspace{0.2cm}

\noindent H8: For AI-generated music, genres typically associated with human performance (e.g., soul, jazz) are rated higher in humanness, while electronic genres are perceived as more artificial.
\vspace{0.2cm}

\noindent \textit{RQ4: Listener Evaluation Criteria for AI-Generated Music}

What aspects of AI-generated do people consider when evaluating it? Current study settings \citep[e.g.,][]{HONG2022107239, moura_artificial_2021} rely on rating scales adapted from adjacent domains. These measures may lack specificity in capturing how people actually experience and judge AI-generated music. Using an exploratory approach, we aim to identify the subjective criteria and themes underlying participants’ judgments, based on open-ended responses.

\section{Methods}

\subsection{Stimuli}
Stimuli were created using the AI tools Stable Audio (v.2.0) and Suno (v.3.5). To create ecologically valid stimuli, prompts were based on findings from the Emotion-to-Music Atlas \citep[EMMA;][]{strauss_emotion--music_2024}, a comprehensive database of human emotion annotations for music excerpts from various genres. In particular, we drew on recent data providing GEMS-9 \citep{zentner_emotions_2008} annotations for 453 music excerpts from the Music4All-Onion database \citep{moscati_music4all-onion_2022}. Excerpts were derived from five major genres with three distinct subgenres each: (1) Electronic (Trance, Downtempo, Dubstep), (2) Jazz (Bossa-Nova, Smooth-Jazz, Swing), (3) Pop (Europop, Indie-Pop, Pop-Punk), (4) Rock (Soft-Rock, Rock-Ballad, Doom-Metal), and Soul/Funk (Funk, Disco, Soul).

For each subgenre, prompts were constructed based on the most prominent emotion dimensions, determined by the mean values across all songs within that subgenre in the EMMA, and their respective GEMS-9 descriptors (e.g., “a joyful, bouncy, animated trance song” to represent the Joyful Activation dimension for the Trance subgenre). This procedure yielded 2-3 prompts per subgenre, resulting in a total of 38 different prompts. These were adapted for use in Suno and Stable Audio (for the full prompt list, see Supplements A), resulting in a set of 76 stimuli.

To account for the properties of each tool, the following approach was used: For Suno, prompts were created in instrumental mode, and one of two outputs was randomly selected and cut to a duration of 45–60 seconds to ensure comparability with prior research \citep[e.g,][]{strauss_emotion--music_2024, warrenburg_choosing_2020}. For Stable Audio, the term ‘instrumental’ was added to each prompt and the target duration set to 60 seconds. Further details on the generation and editing procedure as well as the audio files are provided in the supplementary materials.

\subsection{Sample}

A total of 178 participants took part in the online experiment. Participants were recruited via the University of Innsbruck's mailing list as well as the crowd-sourcing platform Prolific. As an incentive for participation, all participants received individual feedback on their self-reported music behavior and preferences. Additionally, psychology students from the University of Innsbruck (\textit{n}~=~54, 30.3\%) received course credit for their participation, whereas participants recruited via Prolific (\textit{n}~=~124, 69.7\%) received 6 GBP. Participation via Prolific was restricted to subjects reporting English or German as their first language.
About half of the participants took the assessment in German (\textit{n}~=~93, 52.2\%), and half in English (\textit{n}~=~85, 47.8\%). 

An overview of participants’ characteristics can be found in Table~\ref{tab:socio} in the Appendix. No significant differences in person characteristics were found across experimental groups.

\subsection{Procedure and experimental design}

Participants were randomly assigned to one of three groups: (1) In the first group, participants were told that the music excerpts they are going to hear were composed by AI; (2) in the second group, participants were told that the music excerpts are part of movie soundtracks; (3) in the third group, participants were instructed that they will hear music composed by both humans and AI and have to indicate the composer identity for each song. Participants assigned to groups 1 and 3 hence were presented with definitions of AI and AI music tools before the music rating.

After receiving the group-specific instructions, participants were tasked with rating 18 music excerpts being randomly drawn balanced across subgenres and AI tool. Before hearing the first song, participants were instructed to set the volume to a comfortable level and familiarize themselves with all the emotion terms. The order of terms varied randomly between participants but remained constant across songs for each individual. Participants were further instructed to rate the emotions they actually experienced, rather than those they perceived the music to convey. Each music excerpt was presented twice. During the first presentation, participants were instructed to listen attentively without responding. Subsequently, the rating questions were displayed, and the music excerpt was repeated to minimize potential memory-related biases in participants' evaluations.
After completing the music ratings, participants proceeded to the questionnaire part, where they provided information on their personality traits, music-related variables, and socio-demographic background. Finally, participants’ attitudes toward AI were assessed. After completing the questionnaires, participants received a debrief screen which included a description of the study and how the data will be used. Participants were invited to share any additional thoughts on AI-generated music as well as their overall experience of the study through an open-ended question. The survey concluded with the presentation of individual feedback to participants. 

\begin{figure}
    \centering
    \includegraphics[width=1\linewidth]{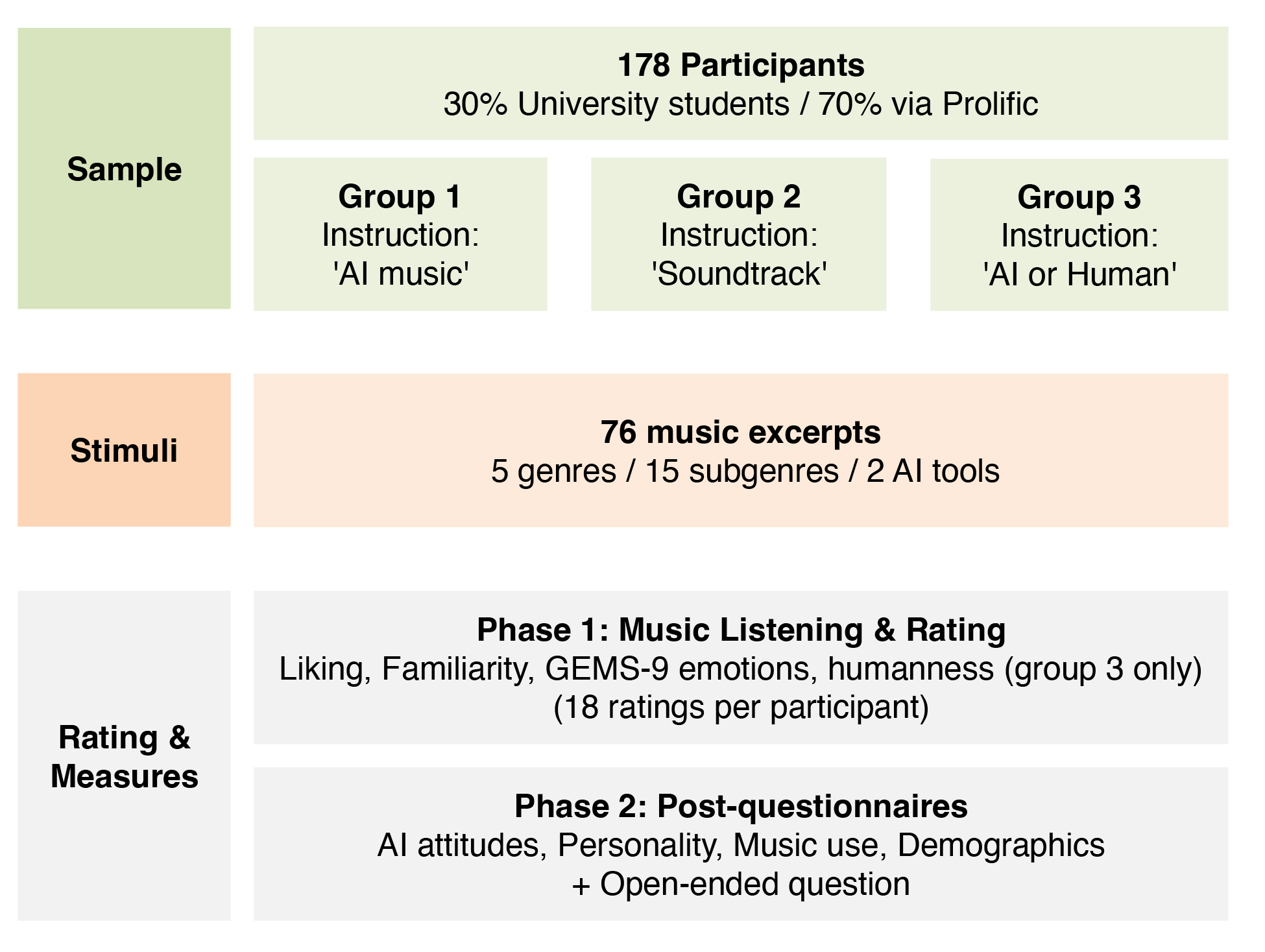}
    \caption{Study Design}
    \label{fig:study_design}
\end{figure}

To ensure data quality, the survey included an attention check similar to the one used by \cite{shank_ai_2023} (“Who was the composer of the music you just heard?”) before the music rating, as well as a self-report on answer honesty and care (5-point scale), and a question about the listening setup (headphones vs. speakers) afterwards. Participants were informed that their responses would not affect feedback or compensation. An overview of the study design is presented in Figure~\ref{fig:study_design}.

\subsection{Measures}
\subsubsection{Stimuli rating}
All participants were asked to indicate how well they knew the music excerpt and how much they liked it on scale from 1 (not at all) to 5 (very). If participants selected a value of 4 or higher for familiarity, they were shown an additional yes-or-no-question inquiring if they associated a particular experience or person with that music excerpt. Participants of the “AI or human” condition were additionally asked to indicate whether they thought that the music excerpt was generated by an AI or a human including the confidence level of their response on a 6-point scale, ranging from 1 (definitely AI) to 6 (definitely human). Participants’ emotional responses to the music excerpts were assessed using the Geneva Emotion Music scale \citep[GEMS;][]{zentner_emotions_2008} for all three conditions. The GEMS comprises a total of nine music-specific emotion dimensions that can be summarized in three overarching factors: (1) Sublimity (Wonder, Transcendence, Tenderness, Peacefulness, and Nostalgia), (2) Vitality (Joyful Activation and Power), and (3) Unease (Sadness and Tension). Here, we used the checklist version GEMS-9, which represents the nine GEMS dimensions with one item (i.e., the label of the respective dimension) and 2-3 exploratory terms each. Emotion ratings were provided for each dimension on visual analog scales ranging from 0 to 100, with all initial values being 0. If no answer was chosen, the answer was hence coded as 0. An open text box allowed for adding up to three additional emotion terms not included in the GEMS, plus the respective intensity ratings.

\subsubsection{AI-related aspects}
\textbf{Attitude toward artificial intelligence.} General attitude toward AI was assessed using the ATTARI-12 \citep{stein_attitudes_2024}, which comprises 12 items on cognition, emotion, and behavior toward artificially intelligent technology. Answers were provided on a scale from 1 (strongly disagree) to 5 (strongly agree). The scale was originally validated in both English and German \citep{stein_attitudes_2024}. Internal consistencies for ATTARI subscales were good, with $\omega$~=~.84--.89, and excellent for the total scale, with $\omega$~=~.95.

At the conclusion of the study, participants were invited to share any additional thoughts on AI-generated music (see Supplements C for detailed phrasing).

\subsubsection{Music background and personality variables}
\textbf{Music competence and appreciation.} The Music-Mindedness Questionnaire \citep[MMQ;][]{zentner_assessing_2017} comprises a four-item Music Competence scale (MMQ-C; e.g., “I can tell when an instrument is out of tune”) and a four-item Music Appreciation scale (MMQ-A; e.g., “Musical experiences are among the most precious in my life”), both rated on a scale from 1 (not at all) to 5 (very much). Internal consistency was $\omega$~=~.90 for MMQ-C and $\omega$~=~.88 for MMQ-A.

\textbf{Motives for music listening.} Participants’ motives for listening to music were assessed by the Uses of Music Inventory \citep[UMI;][]{chamorro-premuzic_personality_2007}. It comprises three dimensions, measured by five items each: (1) Emotional Use (e.g., ‘Listening to music really affects my mood’), (2) Cognitive Use (e.g., ‘I often enjoy analysing complex musical compositions’), and (3) Background Use (e.g., ‘I enjoy listening to music while I work’). Responses are given on a scale from 1 (strongly disagree) to 5 (strongly agree). The items were translated into German and back-translated by three independent experts for an earlier study \citep{strauss_emotion--music_2024}. Internal consistencies were ranging from $\omega$~=~.69-.82.

\textbf{Music genre preferences.} The Short Test of Music Preferences \citep[STOMP;][]{rentfrow_re_2003} assesses participants’ basic preference levels for 14 music genres on a 7-point scale ranging from strongly dislike (1) to strongly like (7). Items were translated into German by two independent experts for an earlier study \citep{strauss_emotion--music_2024}. In the current analyses, we used the four higher level music preference dimensions: Reflective \& Complex, Intense \& Rebellious, Upbeat \& Conventional, and Energetic \& Rhythmic.  Internal consistencies were relatively low, with $\omega$~=~.47-.69.

\textbf{Personality traits.} Participants' personality traits were assessed by using a short version of the Big Five Inventory \citep[BFI-10;][]{rammstedt_measuring_2007}, which was validated in both English and German. The five personality traits Openness for Experience, Consciousness, Extraversion, Agreeableness, and Neuroticism are measured using two items each, and responses are collected on a 5-point scale ranging from 1 (disagree strongly) to 5 (agree strongly). Pearson correlations for Openness for Experience and Agreeableness were small, with \textit{r}~=~.12 and \textit{r}~=~.19 respectively, but large for the other indicators, with \textit{r}s~=~.47.

\textbf{Socio-demographic information.} Basic socio-demographic characteristics were assessed to better understand our sample and examine to what extent they influence the study outcomes. Variables assessed include participants age, gender, nationality, and educational attainment. Although the survey allowed for non-binary responses, all participants identified as male or female. Therefore gender was coded as a binary variable. Finally, language proficiency in the survey language is assessed to ensure that participants can understand subtle differences in adjectives describing their emotional experience.

\subsection{Data analysis}
All statistical analyses were performed in Python (v3.9.20) using the pandas (v2.2.2) library for data handling, scipy (v1.12.0) for correlation and ANOVA testing and statsmodels (v.0.14.4) for mixed-effects models. Figures were created in R (v4.3.3) using the ggplot2 package \citep{ggplot2}.

Thematic analysis, following the methodology by \cite{braun_using_2006}, of open answers to the question about AI-generated music was conducted using MAXQDA (v.24.10). 

\section{Results}
\subsection{Liking, Emotional Perception and Composer Bias}

\begin{figure}
    \centering
    \includegraphics[width=1\linewidth]{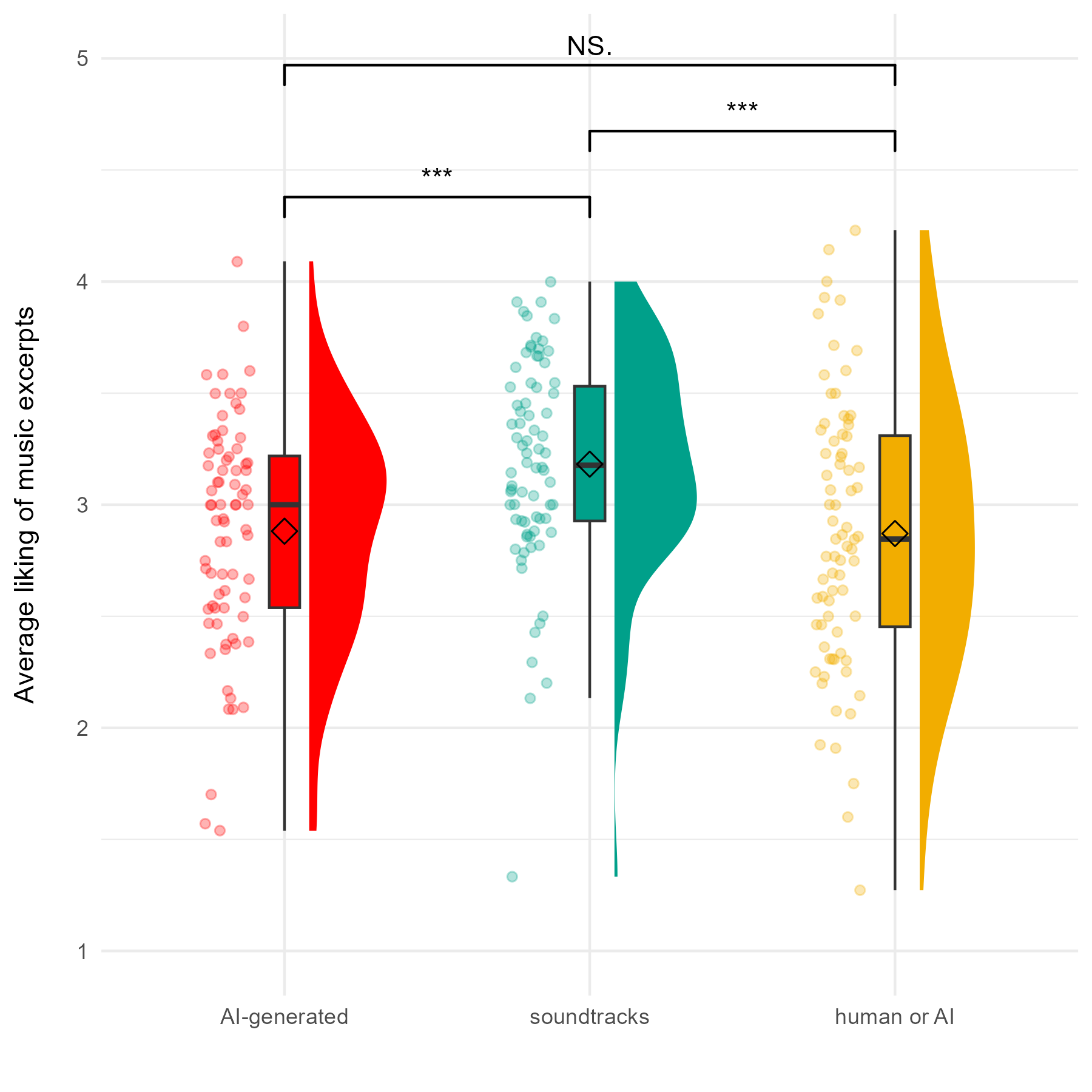}
    \caption{Average Liking per experimental group}
    \label{fig:boxplot_like}
\end{figure}

We examined whether information about a song's origin influenced participants' liking ratings and emotional perceptions. To isolate composer bias effects from song-specific quality differences, we conducted song-level analyses by aggregating participant ratings for each of the 76 songs within the three experimental groups (\textit{n}~=~228 song-group combinations). This approach tests whether identical musical content receives different evaluations based on composer information. A one-way ANOVA revealed significant differences in song-level liking ratings between experimental groups (F(2, 225)~=~8.36, \textit{p}~<~.001, see Figure~\ref{fig:boxplot_like}). Songs labeled as soundtracks ($\mu$~=~3.18, $\sigma$~=~0.47) were significantly rated higher than AI-labeled songs ($\mu$~=~2.88, $\sigma$~=~0.50; \textit{t}~=~-3.79, \textit{p}~<~.001) and unlabeled songs ($\mu$~=~2.87, $\sigma$~=~0.61; \textit{t}~=~3.51, \textit{p}~<~.001). No significant differences emerged between AI-labeled and unlabeled conditions (\textit{t}~=~0.12, \textit{p}~=~.91). Because informing listeners that songs were AI-generated did lower liking relative to framing the same songs as human-composed, H1 was supported.

\begin{figure}
    \centering
    \includegraphics[width=1\linewidth]{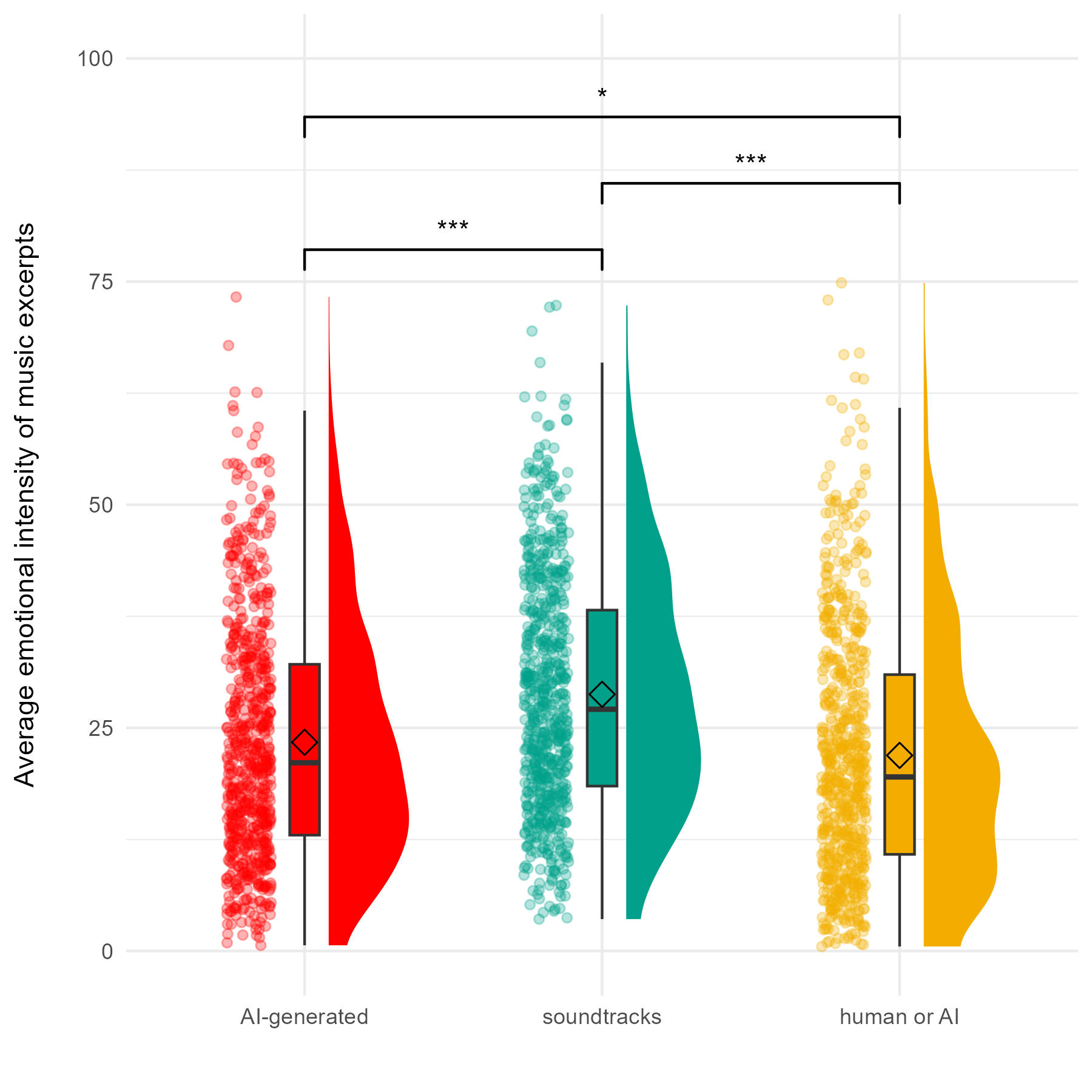}
    \caption{Average emotional intensity per experimental group}
    \label{fig:boxplot_emo}
\end{figure}

Song-level analysis of emotional intensity ratings revealed significant differences between experimental groups (F(2, 225)~=~35.43, \textit{p}~<~.001, Figure~\ref{fig:boxplot_emo}). Songs labeled as soundtracks were perceived as significantly more emotionally intense ($\mu$~=~28.76, $\sigma$~=~5.05) compared to both AI-labeled songs ($\mu$~=~23.38, $\sigma$~=~5.28; \textit{t}~=~-6.41, \textit{p}~<~.001) and unlabeled songs ($\mu$~=~21.91, $\sigma$~=~5.48; \textit{t}~=~8.00, \textit{p}~<~.001). AI-labeled songs showed a trend toward higher emotional intensity ratings than unlabeled songs, though this difference was not statistically significant (\textit{t}~=~1.68, \textit{p}~=~.095).  Since perceived emotional intensity of AI-labeled songs is lower compared to the songs labeled as soundtracks, H2 was supported. 

Additional participant-level robustness analyses, averaging ratings across all songs per participant, are reported in Appendix C. These analyses replicated the composer bias effect for liking but did not reveal significant group differences for emotional intensity.

We tested H3 (whether composer bias effects varied across music genres) using factorial ANOVAs with genre and experimental group interactions for liking and emotional intensity ratings. No significant interactions were found for either liking (F(28, 183)~=~0.29, \textit{p}~>~.999) or emotional intensity (F(28, 183)~=~0.28, \textit{p}~>~.999). Aggregating the 15 subgenres into main genres (Electro, Pop, Rock, Funk/Soul, Jazz) led to the same result: no interaction effects for liking (F(8, 213)~=~0.16, \textit{p}~=~.996) or emotional intensity (F(8, 213)~=~0.17, \textit{p}~=~.995). Therefore, H3 was rejected.

\subsection{The influence of attitudes, personal characteristics and music preferences on liking and perceived emotional intensity of AI-generated music}

To analyze influences of attitudes and personal characteristics, we averaged the liking ratings and the emotional intensity ratings of each participant over all 18 songs they rated. A strong positive correlation was found between participants’ overall attitudes toward AI and their average liking of the AI-generated music (\textit{r}~=~.37, \textit{p}~<~.001). This pattern held when breaking down attitudes into their affective (\textit{r}~=~.32) cognitive (belief-based; \textit{r}~=~.31), and behavioral (intent-to-use; \textit{r}~=~.40) components (all \textit{p}~<~.001), with each showing significant positive associations with liking. Similarly, overall AI attitudes correlated positively with emotional intensity ratings (\textit{r}~=~.31, \textit{p}~<~.001), and again this relationship was consistent across affective, cognitive, and behavioral attitude dimensions (\textit{r}~=~.25, .28, and .34, respectively; all \textit{p}~<~.001). These findings support H4.

Several personality and music-related variables correlated with both liking and emotional intensity (e.g., conscientiousness and reflective/conventional preferences positively; neuroticism negatively; selected motives and competence positively; see Supplements table S2 for full coefficients). When modeled jointly in a linear regression, only attitudes toward AI remained a robust positive predictor of liking ($\beta$~=~0.18, \textit{p}~=~.002) when accounting for individual differences. Among the other personality and music related variables, only neuroticism was a significant predictor, showing a negative association with liking ($\beta$~=~−0.12, \textit{p}~=~.032). All other predictors did not contribute significantly within this multivariate context. Accordingly, we controlled for the variables that correlated with average emotional intensity. Again, only overall attitudes toward AI significantly predicted stronger emotional responses to the music ($\beta$~=~3.06, \textit{p}~=~.019); none of the personality or music-related control variables showed significant effects. Hence, H5 was not supported, apart from a minor negative effect of neuroticism.

\subsubsection{Reassessing composer bias}

To assess whether experimental condition influenced liking and emotional intensity ratings beyond individual differences and music stimuli, we fit linear mixed-effects models with crossed random intercepts for participants and songs while controlling for personality, listening characteristics and the music model used. The results are presented in Table~\ref{tab:lmm_liking_emoint}.

\begin{table*}[t]
\centering
\begin{threeparttable}
\caption{Multilevel Regression Results: Predictors of Liking and Emotional Intensity in Music Excerpts}
\label{tab:lmm_liking_emoint}
\footnotesize
\renewcommand{\arraystretch}{1.15}

\begin{tabular*}{\textwidth}{@{\extracolsep{\fill}} l l c c c c c c}
\cmidrule(lr){1-8}
 &  & \multicolumn{3}{c}{\textbf{Liking}} & \multicolumn{3}{c}{\textbf{Emotional intensity}} \\
\cmidrule(lr){3-5} \cmidrule(lr){6-8}
\textbf{Predictor} & \textbf{Level} & $\beta$ & 95\% CI & $p$ & $\beta$ & 95\% CI & $p$ \\
\cmidrule(lr){1-8}
Intercept &  & 1.43 & [0.34, 2.52] & .010 & 1.34 & [-21.33, 24.00] & .908 \\
\addlinespace[0.3em]
\multicolumn{8}{@{}l}{\textit{Experimental group (ref: AI)}} \\
\addlinespace[0.15em]
 & Soundtracks & 0.21 & [-0.02, 0.43] & .071 & 3.49 & [-1.20, 8.17] & .145 \\
 & Human or AI & -0.04 & [-0.27, 0.20] & .769 & -1.07 & [-5.90, 3.75] & .662 \\
\addlinespace[0.3em]
\multicolumn{8}{@{}l}{\textit{GenAI model (ref: Stable Audio)}} \\
\addlinespace[0.15em]
 & Suno & 0.59 & [0.52, 0.66] & $<$.001 & 4.44 & [3.77, 5.10] & $<$.001 \\
\addlinespace[0.3em]
\multicolumn{8}{@{}l}{\textit{Controls}} \\
\addlinespace[0.15em]
Gender & Female & -0.07 & [-0.27, 0.12] & .463 & -1.56 & [-5.60, 2.49] & .452 \\
Age &  & 0.00 & [-0.01, 0.00] & .469 & -0.03 & [-0.18, 0.11] & .670 \\
\addlinespace[0.3em]
ATTARI Total score &  & 0.11 & [0.04, 0.27] & .010 & 0.15 & [0.69, 5.50] & .012 \\
\addlinespace[0.3em]
BFI Openness for experience &  & -0.04 & [-0.19, 0.06] & .319 & -0.21 & [-7.12, -2.02] & $<$.001 \\
BFI Conscientiousness &  & 0.03 & [-0.08, 0.15] & .537 & 0.10 & [-0.55, 4.33] & .129 \\
BFI Extraversion &  & -0.05 & [-0.16, 0.05] & .285 & -0.06 & [-3.18, 1.10] & .340 \\
BFI Agreeableness &  & 0.04 & [-0.06, 0.17] & .362 & 0.05 & [-1.45, 3.47] & .420 \\
BFI Neuroticism &  & -0.12 & [-0.26, -0.02] & .022 & -0.11 & [-4.44, 0.57] & .129 \\
\addlinespace[0.3em]
UMI Emotional use &  & 0.09 & [-0.03, 0.34] & .106 & 0.02 & [-3.46, 4.35] & .823 \\
UMI Cognitive use &  & -0.03 & [-0.20, 0.11] & .570 & 0.11 & [-0.84, 5.60] & .147 \\
UMI Background use &  & -0.01 & [-0.20, 0.15] & .796 & -0.03 & [-4.39, 2.86] & .679 \\
STOMP Reflective \& Complex &  & 0.05 & [-0.05, 0.15] & .358 & 0.09 & [-0.76, 3.36] & .216 \\
STOMP Intense \& Rebellious &  & -0.01 & [-0.08, 0.06] & .803 & -0.06 & [-2.17, 0.65] & .291 \\
STOMP Upbeat \& Conventional &  & 0.08 & [-0.01, 0.18] & .097 & 0.04 & [-1.40, 2.56] & .565 \\
STOMP Energetic \& Rhythmic &  & 0.03 & [-0.05, 0.10] & .551 & 0.09 & [-0.44, 2.78] & .154 \\
MMQ Music competence &  & 0.03 & [-0.10, 0.18] & .561 & 0.08 & [-1.46, 4.26] & .338 \\
MMQ Music appreciation &  & 0.03 & [-0.13, 0.21] & .635 & 0.10 & [-1.46, 5.50] & .256 \\
\cmidrule(lr){1-8}
\end{tabular*}
\begin{tablenotes}
\footnotesize
\item CI~=~confidence interval. 
\end{tablenotes}
\end{threeparttable}
\end{table*}

For liking ratings, experimental condition effects were non-significant when controlling for individual differences in AI attitudes and other personality factors. AI attitudes remained a significant predictor (\(\beta~=~0.152\), \textit{p}~=~.010), while experimental conditions showed no meaningful effects: "Soundtracks (human)" versus AI was non-significant (\(\beta~=~0.207\), \textit{p}~=~.071), and the unlabeled condition versus AI was also non-significant (\(\beta~=~-0.035\), \textit{p}~=~.769). Additionally, neuroticism emerged as a significant predictor, with higher levels associated with lower liking ratings (\(\beta~=~-0.140\), \textit{p}~=~.022). Music stimuli showed a strong effect, with Suno-generated tracks rated significantly higher than Stable Audio tracks ($\beta$~=~0.591, \textit{p}~<~.001). 

For emotional intensity ratings, a parallel analysis confirmed these findings. When controlling for individual AI attitudes and personality factors using crossed random effects, experimental condition effects were non-significant. AI attitudes remained a highly significant predictor (\(\beta~=~3.097\), \textit{p}~=~.012). Neither the "Soundtracks (human)" condition (\(\beta ~=~3.485\), \textit{p}~=~.145) nor the "unlabeled" condition (\(\beta ~=~-1.074\), \textit{p}~=~.662) differed significantly from the AI condition. Openness to experience showed a significant negative association with emotional intensity ratings (\(\beta ~=~-4.568\), \( \textit{p}~<~.001 \)), suggesting that more open individuals perceived the music as less emotionally intense. Again, music stimuli showed an effect, with Suno-generated tracks perceived as significantly more emotionally intense than Stable Audio tracks (\(\beta ~=~4.437\), \( \textit{p}~<~.001 \)).

To refine these results, we repeated the analyses using the ATTARI subscales. Only the behavioral subscale (BEH) emerged as a significant predictor, whereas the affective and cognitive subscales did not contribute once entered simultaneously (see supplements S3). This suggests that the predictive power of AI attitudes is primarily carried by participants’ willingness to engage with AI in practice, rather than by their general beliefs or feelings.

To further examine whether AI attitudes moderated experimental effects, we conducted moderation analyses testing interactions between experimental conditions and both the total ATTARI score and the BEH subscale. For both liking and emotional intensity, no significant moderation effects emerged (all interaction p-values \(~<~.001 \); see supplements S4). Thus, AI attitudes predicted evaluations consistently across all experimental groups, rather than moderating condition effects.

\subsection{Perceived Humanness of AI-generated music}

\begin{figure}[t]
    \centering
    \includegraphics[width=1\linewidth]{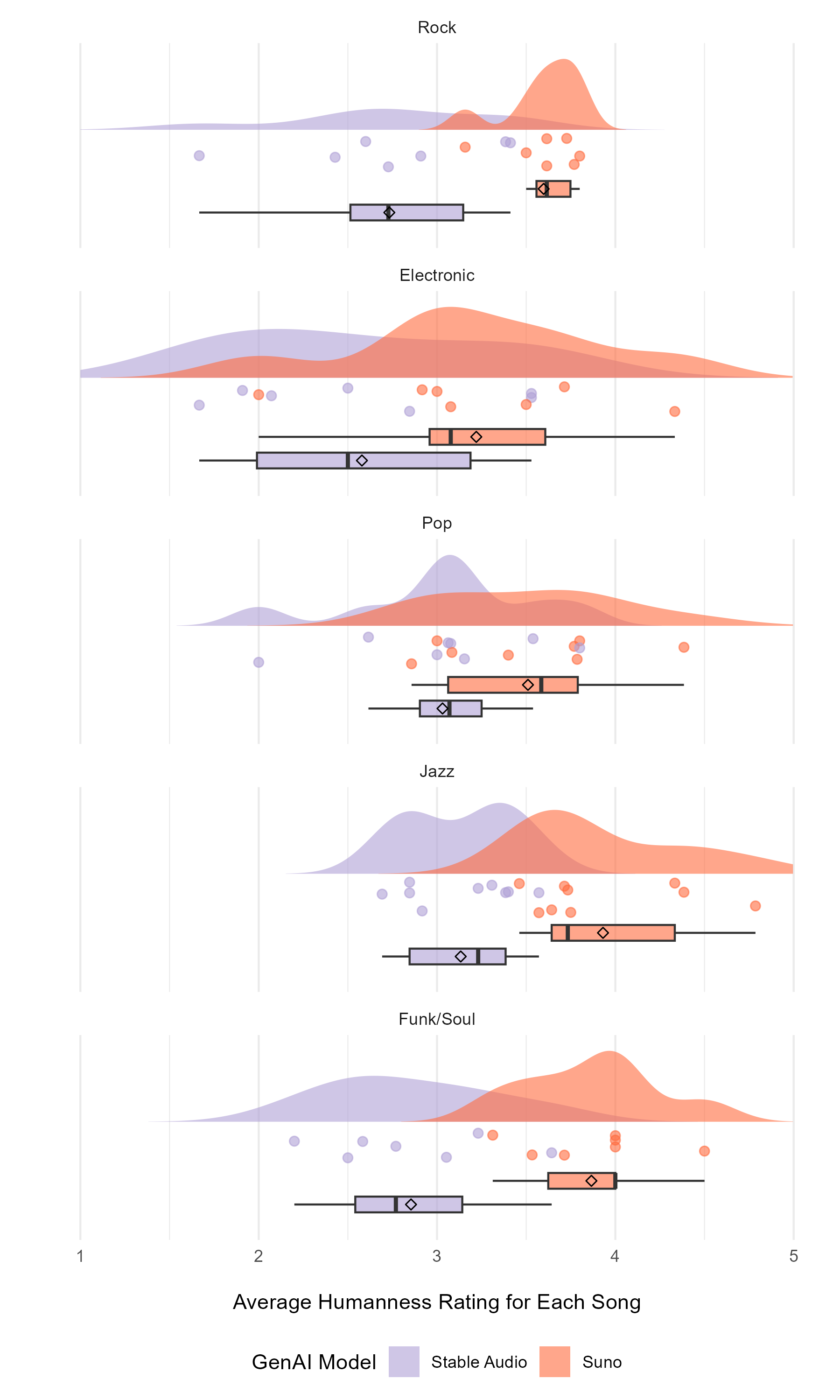}
    \caption{Humanness rating by main genre and stimulus}
    \label{fig:humanness_rating_genre}
\end{figure}

To test H6 (that AI-generated music is liked more when perceived as human-created, regardless of composer identity), analyses were restricted to ratings in the experimental group that received no information about the composer’s identity (1026 ratings for \textit{n}~=~76 songs, $\mu$~=~3.26, $\sigma$~=~.65). We averaged both liking and humanness ratings for each song, and a highly significant correlation was found (\textit{r}~=~0.78, \( \textit{p}~<~.001 \)).

Higher humanness ratings were also correlated with higher average emotional intensity ratings for a song (\textit{r}~=~0.58, \( \textit{p}~<~.001 \)). In more detail, perceived humanness of a track was positively correlated with emotional intensity of sublimity (\textit{r}~=~.61, \( \textit{p}~<~.001 \)), and negatively correlated with feelings of unease (\textit{r}~=~-.43, \( \textit{p}~<~.001 \)). There was no association between humanness ratings and the vitality dimension. Thus, H6 and H7 are supported.

To examine differences in perceived humanness between AI music generation 
models, we conducted an independent samples t-test comparing mean humanness ratings at the song level (\textit{n}~=~76 songs). Results showed that Suno-generated music ($\mu$~=~3.64, $\sigma$~=~0.53, \textit{n}~=~38) was rated significantly higher in humanness than Stable Audio-generated music ($\mu$~=~2.88, $\sigma$~=~0.56, \textit{n}~=~38), t(74)~=~6.04, \( \textit{p}~<~.001 \), Cohen's d~=~1.39. 

Genres rated as most human included swing ($\mu$~=~3.74), soul ($\mu$~=~3.60), and soft-rock ($\mu$~=~3.48) while dubstep ($\mu$~=~2.54) pop-punk ($\mu$~=~2.69), and trance ($\mu$~=~2.74) received the lowest humanness ratings. A Kruskal-Wallis-Test showed no significant differences in humanness ratings between genres on song level. However, when looking at the songs of each AI-model separately, a Kruskal-Wallis test for songs generated with Suno showed significant differences betweeen genres (H~=~24.97, \textit{p}~=~0.0348, \textit{n}~=~38 songs). An overview of humanness ratings across main genres and stimuli is presented in Figure~\ref{fig:humanness_rating_genre}. Thus, H8 was not supported at the aggregate level, receiving only model-specific, partial support within Suno’s outputs.

\subsection{Listener Evaluation Criteria for AI-Generated Music}

To complement the quantitative analyses and to uncover subjective evaluation criteria of participants, we conducted a thematic analysis of the 178 open-ended responses. An overview of the main themes is presented in Table~\ref{tab:themes}; the full codebook is available in the Supplementary Materials. Categories such as \textit{music quality}, \textit{emotional aspects}, and \textit{humanness} also emerged in the qualitative material. Participants frequently described AI-generated music in terms of its technical and musical quality, its originality, and its ability to evoke emotions. Humanness, the most frequent theme, was not only linked to perceived sound characteristics. Several respondents emphasized the importance of the human touch, that the personality of an artist, the history behind a song influenced their judgements, and described AI-generated music as having no soul and authenticity. At the same time, some participants expressed indifference to the song's origin ("If it sounds good then I don't mind if it was artificially generated", ID926).

\begin{table}[t]
\caption{Themes from the thematic analysis of open-ended responses}
\label{tab:themes}
\renewcommand{\arraystretch}{1.15}
\small
\begin{tabularx}{\columnwidth}{p{0.35\columnwidth} X}
\toprule
\textbf{Theme} & \textbf{Description} \\
\midrule
Music Quality & Judgments of technical and musical quality of AI music. \\
Emotional Aspects & Emotional reactions to AI music, different types of emotions and emotional depth. \\
Humanness & Perceptions of authenticity, soul, and human touch, and whether the production process is attributed to humans or AI. \\
Attitudes toward AI & General stances toward AI in art, normative statements, rejection of AI music. \\
Ethical Considerations & References to replacement of artists, economic implications, copyright or training data. \\
Cultural / Artistic \newline Implications & Concerns about loss of creativity or artistry, reflections on changes to the artistic process. \\
Innovation \& \newline Opportunities & Expressions of technological fascination, democratization of music-making, new possibilities for artists. \\
Practical Use / \newline Contextual Value & Situational evaluations of AI music, such as functional music use, personalization, productivity in music-making. \\
\bottomrule
\end{tabularx}
\caption*{\footnotesize Note. The full codebook with subcodes and definitions is available in the Supplementary Materials.}
\end{table}


The open-ended responses provided further evidence that \textit{attitudes toward AI} strongly shape how listeners evaluate AI-generated music. Many comments were explicitly normative and didn't reference quality or characteristics of the songs, for instance stating that they “don’t like anything to do with AI-generated music” (ID914). 

Beyond these dimensions, the open-ended answers highlighted several additional areas of evaluation that go beyond typical measures in current experimental work. First, participants evaluated AI music according to \textit{ethical criteria}. Concerns included the replacement of musicians, potential economic harm to artists, and transparency of training data and copyright ("is built upon data without the consent of the artists that have dedicated decades of their lives to the craft", ID615). For example, one participant noted that "I think it undermines and abuses artists" (ID446), reflecting that listeners not only assess the aesthetic qualities of AI music but also the fairness of its production and distribution.  

Second, many respondents referred to the \textit{cultural implications} of AI in music. Negative perspectives centered on a perceived loss of creativity, oversaturation of low-quality content, and the risk that music becomes “slop” due to low entry barriers. At the same time, some participants saw positive or ambivalent aspects, such as new possibilities for collaboration or co-creativity between human and AI, and valued new skills required to use AI tools effectively ("I have found that using AI generated music is something that requires talent and i appreciate it when i come across it", ID913).  

In addition, a prominent theme concerned \textit{innovation and opportunities}. Many listeners expressed fascination with the capabilities of AI systems and described them as impressive. Others pointed to the democratization of music production ("can open the door for people like me who love music but are not talented in singing or playing instruments", ID610) and to new opportunities for artists, such as inspiration for new genres or creative directions ("gives artist inspirations and abilities to explore new styles", ID46). 

Finally, participants also referred to the \textit{functional use and contextual value} of AI music. Several comments highlighted that it could serve well as functional background, e.g., for studying or relaxation (ID53), for therapy, or as personalized music tailored to individual moods and tastes (ID211). Others pointed to its potential for increasing efficiency in music-making, noting that AI can speed up production processes (ID613). These statements underline that evaluations of AI music are strongly context-dependent.

\section{Discussion}

\subsection{Summary of main findings}
    
Our findings initially suggested a significant effect of manipulation of composer identity for both liking and emotional perception. Tracks labeled as soundtracks (indicating human origin) were significantly rated higher than those labeled AI-generated or labeled as "AI or human".

However, further analysis revealed that this effect was largely driven by individual differences in attitudes toward AI, rather than the experimental manipulation alone. When attitudes toward AI were included as covariates in linear mixed effect models, the experimental condition no longer predicted liking or emotional intensity, whereas attitudes toward AI remained a consistent and significant predictor. This supports previous findings considering attitudes toward autonomous machines predicts evaluation \citep{HONG2022107239} and adds perspective to \cite{shank_ai_2023}'s work suggesting that composer bias might be song and context specific (e.g., present for well-liked, human-sounding songs). 

Although several personality traits and music-related variables correlated with liking and emotional experience, their impact largely diminished when controlled for attitudes toward AI. Notable exceptions are openness to experience and neuroticism, two personality traits that have been previously found to substantially affect the experience of music, particularly in respect to emotions related to joy and vitality \citep{gerstgrasser_role_2023, juslin_emotional_2011, liljestrom_experimental_2013}, both of which were highly prominent in the stimuli of the present study. Most interestingly, attitudes toward AI was the only listener characteristic that was significantly associated with both liking and emotional experience. The total amount of variance explained by listener characteristics thereby was of similar size to that reported in previous research \citep{gerstgrasser_role_2023, strauss_emotion--music_2024}. These findings emphasize that attitudes toward AI are a key determinant in how AI-generated music is evaluated across diverse users. In contrast to \citet{ansani_ai_2024}, we did not find a moderation effect of attitudes toward AI on experimental groups, which may be due to differences in the attitude scales used, our focus on listening to tracks rather than live performance, and the inclusion of three instead of two experimental conditions.

An open question remains: Why did attitudes toward AI predict liking and emotional responses even when composer information was absent or ambiguous (No-Information condition)? A plausible reading is that ATTARI may index a broader technological openness that colors music evaluations even without explicit provenance cues. An alternative is that listeners implicitly inferred AI involvement from sonic characteristics or prior exposure, therefore reintroducing provenance via perception rather than label.

Consistent with earlier studies, we find that higher perceived humanness is related to higher liking \citep{ansani_ai_2024, deguernel_investigating_2022, shank_ai_2023}; we extend this literature by demonstrating that the effect generalizes to emotional intensity. Humanness thus functions as a cue for realism or authenticity, serving as an implicit quality signal.
This is further supported by our qualitative analysis of open-ended questions: “humanness” was the most frequent theme, but invoked not only for sonic traits but for a human touch tied to artist/persona and backstory;

Importantly, many remarks were normative about AI itself rather than the audio, reinforcing that attitudes toward AI drive evaluations even without explicit composer information. 
Listeners raised concerns that mirror debates in the music information retrieval and generative audio research communities, including potential negative economic implications, ethical issues of authorship and copyright, and broader cultural consequences \citep{holzapfel_ethical_2018, barnett_ethical_2023}. At the same time, positive contextual factors and technological curiosity also shaped responses. Taken together, this may also help explain why findings on composer bias remain mixed across the literature: perceptions of AI are not static, but shift in line with technological progress. As generative systems become more capable, what once was seen with fascination can increasingly be perceived as a challenge or even a threat. Perceptions of AI-music could therefore be seen as dynamic rather than fixed.

\subsection{Limitations}

The study has several limitations: First, we used a bilingual, online convenience sample consisting of university students and a Prolific sample with uncontrolled listening setups (headphones/speakers, self-set volume), which limits generalizability. Also, as mentioned in previous work on AI-music \citep{shank_ai_2023}, participants recruited via Profilic are more likely to possess technological literacy and prior familiarity with AI. 

Second, framing in one experimental group for the stimuli as "movie soundtracks" could have introduced unintended bias, since people tend to have positive connotations to film music.

Finally, we restricted stimuli to instrumental excerpts and our results may not generalize to vocal music. Previous research indicates that lyrics generally tend to enhance the emotional intensity of sad songs, whereas for happy songs they gave been found to exert little or even diminishing effects \citep{ali_songs_2006,barradas_when_2021, brattico_functional_2011, warmbrodt_emotion_2022}. Given the predominance of positively valenced music in the present study, hence only a marginal (diminishing) effect on emotional intensity would have been expected by the addition of lyrics, assuming congruent lyrics with a quality similar to human composed music. Our choice therefore aimed to minimize confounds from lyrical semantics and prosody and reflected the limited, reliable availability of high-quality vocal generation at the time of stimulus creation.

\subsection{Future Studies}
Our findings about the influence of attitudes, together with the qualitatively identified themes, particularly ethical considerations, functional use, point to music perception beyond aesthetic measures. Future work should adopt instruments specifically tailored to AI music and develop multi-factor rating scales accordingly.

With AI music increasingly present on streaming platforms, socioeconomic questions about fairness and transparency surrounding AI-generated music will have to be examined. Given the influence of attitudes toward AI on music liking and emotions, experimental designs manipulating information about the creation and distribution process could be addressed. For instance, transparency of training data or intended revenue share of AI-music pieces could be modeled as an experimental condition. 

Furthermore, different production and usage contexts could be studied in more detail. For instance, the perception of songs that are created collaboratively between human and AI systems is a growing topic of interest, as current text-to-music systems increasingly enable processes of co-creation. Also, expanding on the work of \cite{moura_artificial_2021}, the perception and acceptance of AI music in functional use cases could be examined.

\section{Conclusion}

Our findings highlight that attitudes toward AI play a central role in shaping how listeners perceive and evaluate AI-generated music, outweighing composer information, genre, and listening habits or other personal characteristics. 

Looking ahead, it will be crucial to observe how general attitudes toward artificial intelligence develop over time, since they are likely to shape both acceptance and resistance in cultural domains such as music. More broadly, the debate is not only about whether AI-generated music can be enjoyed or recognized as human-like, but about what counts as “authentic” and “valuable” creativity. By critically engaging with audience responses, we can address questions like: What kinds of (artificial) creativity do we want to acknowledge, and under which terms?

\section*{CRediT authorship contribution statement}

David Stammer: conceptualization, formal analysis, investigation, project adminsitration, data curation, methodology, visualization, and writing - original draft

Hanna Strauss: conceptualization, investigation, methodology, project administration, validation, data curation, resources, formal analysis, visualization, and writing – original draft.  

Peter Knees: conceptualization, methodology, validation, writing - review \& editing, and supervision.

\section*{Acknowledgements}
The authors are grateful to Peer-Ole Jacobsen for his valuable input on the methodological design of the study.

\section*{Declaration of Competing Interest}
The authors declare that they have no known competing financial interests or personal relationships that could have appeared to influence the work reported in this paper.

\section*{Ethics statement}
The study was conducted in accordance with ‘Ethical Guidelines for Surveys and the Analysis of Online Activities’ which were approved by the Review Board of the Institute of Psychology as well as the Ethics Advisory Board for Scientific Research at the University of Innsbruck (Certificate of good standing, 90/2022).




\appendix

\section{Supplementary Data}
Supplementary material related to this article can be found in the attached file: Supplements.pdf

\section{Socio-demographic data of participants}
\label{app1}

\begin{table*}[t]
\centering
\small
\begin{threeparttable}
\caption{Socio-demographic data of participants by experimental condition and for the total sample (\textit{n}~=~178).}
\label{tab:socio}
\renewcommand{\arraystretch}{1.15}

\begin{tabularx}{\textwidth}{l *{3}{>{\centering\arraybackslash}X} >{\centering\arraybackslash}X}
\toprule
 & \makecell{AI-generated\\(\textit{n}~=~56)} 
 & \makecell{Soundtracks\\(\textit{n}~=~63)} 
 & \makecell{Human or AI\\(\textit{n}~=~59)} 
 & \makecell{Total sample\\(\textit{n}~=~178)} \\
\midrule
\multicolumn{5}{l}{\textit{Gender}} \\
Male   & 28 (50.0\%) & 30 (47.6\%) & 28 (47.5\%) & 86 (48.3\%) \\
Female & 28 (50.0\%) & 33 (52.4\%) & 31 (52.5\%) & 92 (51.7\%) \\
Other  & 0 (0\%)     & 0 (0\%)     & 0 (0\%)     & 0 (0\%) \\
\addlinespace[0.3em]
\multicolumn{5}{l}{\textit{Age (years)}} \\
Mean (SD)            & 33.1 (14.8) & 36.0 (14.5) & 36.2 (14.6) & 35.2 (14.6) \\
Median    & 27.5  & 33.0  & 33.0  & 31.0 \\

[Min, Max]    & [18.0,~75.0] & [20.0,~85.0] & [18.0,~77.0] &  [18.0,~85.0] \\
\addlinespace[0.3em]
\multicolumn{5}{l}{\textit{Highest educational achievement}} \\
Compulsory school without vocational training & 2 (3.6\%) & 2 (3.2\%) & 0 (0\%)   & 4 (2.2\%) \\
Compulsory school with vocational training    & 1 (1.8\%) & 1 (1.6\%) & 1 (1.7\%) & 3 (1.7\%) \\
Vocational school with A-levels/high-school diploma & 5 (8.9\%) & 5 (7.9\%) & 6 (10.2\%) & 16 (9.0\%) \\
Academic High School/Grammar School           & 14 (25.0\%) & 13 (20.6\%) & 16 (27.1\%) & 43 (24.2\%) \\
University/College                            & 34 (60.7\%) & 42 (66.7\%) & 36 (61.0\%) & 112 (62.9\%) \\
\addlinespace[0.3em]
\multicolumn{5}{l}{\textit{Self-reported musicianship status}} \\
Non-musician              & 11 (19.6\%) & 8 (12.7\%) & 7 (11.9\%) & 26 (14.6\%) \\
Music loving non-musician & 28 (50.0\%) & 35 (55.6\%) & 33 (55.9\%) & 96 (53.9\%) \\
Amateur musician          & 15 (26.8\%) & 10 (15.9\%) & 12 (20.3\%) & 37 (20.8\%) \\
Semi-professional musician& 1 (1.8\%)  & 7 (11.1\%)  & 3 (5.1\%)   & 11 (6.2\%)  \\
Professional musician     & 1 (1.8\%)  & 3 (4.8\%)   & 4 (6.8\%)   & 8 (4.5\%)   \\
\addlinespace[0.3em]
\multicolumn{5}{l}{\textit{Amount of music listening}} \\
Never           & 0 (0\%) & 0 (0\%) & 0 (0\%) & 0 (0\%) \\
Sometimes       & 3 (5.4\%) & 4 (6.3\%) & 3 (5.1\%) & 10 (5.6\%) \\
1--2 days/week  & 4 (7.1\%) & 2 (3.2\%) & 3 (5.1\%) & 9 (5.1\%)  \\
3--4 days/week  & 8 (14.3\%) & 5 (7.9\%) & 11 (18.6\%) & 24 (13.5\%) \\
5--6 days/week  & 8 (14.3\%) & 4 (6.3\%) & 8 (13.6\%) & 20 (11.2\%) \\
Every day       & 33 (58.9\%) & 48 (76.2\%) & 34 (57.6\%) & 115 (64.6\%) \\
\bottomrule
\end{tabularx}
\begin{tablenotes}[flushleft]
\item Note. No significant group differences emerged for any socio-demographic variable, all \( \textit{p}~>~.1 \).
\end{tablenotes}
\end{threeparttable}
\end{table*}

\section{Participant-level robustness analyses of composer-bias}

To complement the main song-level analyses reported in the manuscript, we also examined composer bias effects at the participant level by averaging each participant’s ratings across all songs they evaluated.

\subsection*{Liking}
At the participant level, average liking across all songs ($\mu$~=~2.98, $\sigma$~=~0.68) differed significantly between conditions, F(2,175)~=~3.37, \textit{p}~=~.037. Participants rated soundtrack-labeled music ($\mu$~=~3.15, $\sigma$~=~0.57) higher than both AI-labeled music ($\mu$~=~2.87, $\sigma$~=~0.76; t(100.6)~=~-2.28, \textit{p}~=~.025) and unlabeled music($\mu$~=~2.89, $\sigma$~=~0.67; t(113.6)~=~2.30, \textit{p}~=~.024). No difference emerged between AI-labeled and unlabeled conditions t(109.6)~=~-0.18, \textit{p}~=~.86.

\subsection*{Emotional Intensity}
At the participant level, average emotional intensity ratings did not differ significantly between conditions, \(F(2,175)~=~0.27, \textit{p}~=~.767\). Mean ratings were similar across groups (AI-generated: $\mu$~=~29.18, $\sigma$~=~16.34; No-information: $\mu$~=~29.42, $\sigma$~=~14.81; Soundtracks: $\mu$~=~31.02, $\sigma$~=~14.28). Post-hoc Welch tests confirmed the absence of significant pairwise differences (all \textit{p}~>~.50).

\clearpage

\bibliographystyle{elsarticle-harv}

\bibliography{references}

\end{document}